\documentclass[dvipdfmx,preprint]{ptephy_v2}

\preprintnumber{} 
\usepackage{hyperref}
\usepackage{comment}
\usepackage{graphicx}
\usepackage{bm}
\usepackage{amsmath}

\begin{document}

\title{Highly polarized epithermal neutron beam using polarized $^3$He gas at J-PARC}


\author[1]{K. Asai}
\author[2]{S. Endo}
\author[1]{S. Hayashi}
\author[2]{H. Isozaki}
\author[3]{T. Ino}
\author[2]{W. Kambara}
\author[1]{S. Kawamura}
\author[2]{A. Kimura}
\author[1]{M. Kitaguchi}
\author[1]{S. Kudo}
\author[2]{R. Kobayashi}
\author[2]{T. Oku}
\author[1,2]{T. Okudaira*}
\author[1]{M. Okuizumi}
\author[1]{H. M. Shimizu}
\author[1]{K. Shimodaira}
\author[4,2]{Y. Suto}
\author[2]{S. Takada}
\author[5]{S. Takahashi}
\author[2]{H. Tanaka}
\author[2]{Y. Tsuchikawa}
\author[4,2]{T. Yamaguchi}

\affil[1]{Nagoya University, Furocho, Chikusa, Nagoya, Aichi 464-8602, Japan 
\email{okudaira@phi.phys.nagoya-u.ac.jp}}
\affil[2]{Japan Atomic Energy Agency, 2-4 Shirakata, Tokai, Naka, Ibaraki 319-1195, Japan}
\affil[3]{High Energy Accelerator Research Organization, 1-1 Oho, Tsukuba, Ibaraki 305-0801, Japan}
\affil[4]{Ibaraki University, 2-1-1 Bunkyo, Mito, Ibaraki 310-8512, Japan}
\affil[5]{The University of Tokyo, Kashiwa Campus 5-1-5 Kashiwanoha, Kashiwa, Chiba 277-8561, Japan}



\begin{abstract}%
We have realized a highly polarized epithermal neutron beam using an in-situ polarized $^{3}$He spin filter at the high-intensity pulsed neutron facility J-PARC. A neutron polarization exceeding 80\% was achieved for epithermal neutrons with energies below 1 eV, including a polarization of $86.4\pm0.5\%$ for 0.75~eV neutrons, which corresponds to the resonance energy of the $p$-wave compound state in the $^{139}$La+n system. The spin filter consisted of a 20-cm-long, 3.2-atm $^{3}$He gas cell polarized on the beamline using a compact laser-optics system for spin-exchange optical pumping (SEOP).  A $^{3}$He polarization of approximately 78\% was stably maintained on the beamline for more than 100~hours. The polarized epithermal neutron beam developed in this work provides a powerful tool for studies of discrete symmetry violations, including measurements of enhanced parity violation in neutron absorption by medium-heavy nuclei such as $^{139}$La and searches for time-reversal symmetry violation in neutron--nucleus systems.
\end{abstract}

\subjectindex{G12, C03, C34, D02, D50}

\maketitle

\section{Introduction}

Polarized neutron beams are powerful tools in a wide range of scientific fields, including particle and nuclear physics, condensed matter physics, and engineering. In particular, polarized-neutron reactions involving compound-nuclear resonances provide a unique opportunity to study parity (P) and time-reversal (T) symmetry violations in neutron-nucleus interactions. 

In epithermal neutron capture by medium-heavy nuclei, P-violating effect induced by the hadronic weak interaction can be strongly enhanced in compound-nuclear resonances through a mixing between $s$-wave and $p$-wave amplitudes~\cite{Mitchell1999}. The largest example in the eV region is the 9.57$\pm$0.35\% helicity-dependent absorption asymmetry, observed at the 0.75~eV $p$-wave resonance of $^{139}$La+n~\cite{LANL91}. This corresponds to an enhancement of approximately six orders of magnitude relative to P-violation in nucleon--nucleon scattering. Theory suggests that the same mechanism can also enhance T-violating effects in the nucleon-nucleon interaction~\cite{Gudkov1991,gud17}. Recent J-PARC measurements of $(n,\gamma)$ spin correlations~\cite{okuda18,yama20} and spin-dependent neutron--nucleus interactions~\cite{Okudaira2024} were combined to quantify the sensitivity of the 0.75~eV resonance of the $^{139}$La+n in P-odd/T-odd interactions, demonstrating that this resonance can be a highly sensitive probe of the T-violating effect~\cite{Nakabe2024}. The NOPTREX (Neutron Optical Parity and Time Reversal EXperiment) collaboration plans to conduct a sensitive search for the T-violating interactions beyond the standard model by probing P-odd/T-odd terms in the forward scattering amplitude of 0.75~eV polarized neutrons incident on polarized $^{139}$La targets at J-PARC.

To polarize neutrons in this energy region, either a proton spin filter or a $^{3}$He spin filter is available. A proton spin filter utilizes the spin-dependent neutron--proton scattering cross section to polarize a neutron beam by transmitting through a polarized proton target. Its advantage is that the scattering cross section is approximately energy independent, allowing neutron polarization of up to 70\% over a wide energy range from eV to keV~\cite{Lushchikov1970, Ishida1984, Bowman1989}. However, because neutron polarization is achieved through the scattering process, this method produces substantial backgrounds from scattered neutrons and requires strong magnetic fields of several tesla and cryogenic temperatures below 1 K to polarize the proton target using dynamic nuclear polarization (DNP) method. Recently, the polarized proton target using Triplet DNP method, which might allow us to use it as the spin filter with relatively low magnetic field and higher temperature, has been developed~\cite{Takada2020,Quan2019, TATEISHI2026142606}. 

A $^3$He spin filter is a neutron polarization device in which polarized $^3$He gas is sealed in a glass cell.
 This device exploits the strongly spin-dependent neutron absorption cross section of $^{3}$He, enabling the production of a high-quality polarized neutron beam with minimal scattering background. Polarization of $^{3}$He gas can be achieved by spin-exchange optical pumping (SEOP)~\cite{SEOP}  or metastability-exchange optical pumping (MEOP)~\cite{MEOP}, and high polarization can be maintained under low magnetic fields of the order of 1~mT without the need for cryogenic conditions. These features allow for a compact and flexible experimental system and have led to the widespread use of $^3$He spin filters at neutron facilities for polarized-neutron experiments~\cite{Babcock2023, Dhiman2017, Chen2017,  Lee2016, Beecham2011, Zhang2025, Mitchell2026}. However, since the neutron absorption cross section of $^{3}$He decreases as $1/\sqrt{E_n}$, where $E_n$ is the neutron energy, achieving high neutron polarization at higher neutron energies requires a large areal density of highly polarized $^{3}$He gas.

At J-PARC, the world's leading high-intensity pulsed neutron facility, the $^3$He spin filter based on the SEOP method has been employed in various neutron experiments~\cite{Ino2017Polano,He3Okudaira,Takahashi2025}. Previous J-PARC measurements of nuclear reactions induced by polarized neutrons~\cite{yama20, endo22, Endo2024Circular, Okudaira2024} used ex-situ polarized $^{3}$He spin filters, where the cells were polarized at an off-beamline SEOP station before being transported to the beamline. With 6--7.5~cm-long, 3.2~atm $^{3}$He cells, an averaged $^{3}$He polarization of approximately 70\% was achieved during the experiment, yielding neutron polarizations of 30--40\% at 0.75~eV. In this work, we developed a compact in-situ polarized $^{3}$He spin filter system, in which a 20~cm-long, 3.2~atm $^{3}$He cell is polarized on the neutron beamline, and demonstrated the production of a highly polarized neutron beam in eV region.

\section{Design of the $^{3}$He spin filter}
The neutron spin filtering effect of polarized $^{3}$He is based on the strong spin dependence of the neutron absorption reaction
\begin{equation}
n + {}^{3}\mathrm{He} \rightarrow {}^{4}\mathrm{He}^{*} \rightarrow p + t + 764~\rm{keV}.
\end{equation}
Since the intermediate $^{4}$He state has total angular momentum zero, the absorption cross section strongly depends on the relative spin orientation between the incident neutron and the $^{3}$He nucleus. 
The neutron--$^{3}$He total cross section can be written as
\begin{equation}
\sigma(E_n)
=
\sigma_0(E_n)
-
\sigma_p(E_n)\,\bm{s}\cdot\bm{I},
\label{eq:he3_cross_section}
\end{equation}
where $\bm{s}$ and $\bm{I}$ denote unit vectors parallel to the neutron spin and the $^{3}$He nuclear spin, respectively. 
The spin-independent term is given by
\begin{equation}
\sigma_0(E_n)
=
\sigma_{\rm{sc}}
+
\sigma_a(E_n),
\end{equation}
where $\sigma_{\rm{sc}}$  and $\sigma_a(E_n)$ are the scattering cross section and the absorption cross section for the unpolarized state.  
The spin-dependent term consists of the spin-dependent scattering and absorption components:
\begin{equation}
\sigma_p(E_n)
=
\sigma_{\mathrm{sc}, p}
+
\sigma_a(E_n).
\end{equation}
The spin-independent and spin-dependent scattering cross sections are $\sigma_{\rm{sc}} = 2.7~\mathrm{barn}$ and $\sigma_{\mathrm{sc},p}=1.4~\mathrm{barn}$, respectively~\cite{Zimmer2002}. The scattering cross sections are treated as energy independent in the neutron-energy region considered here, whereas the absorption cross section follows the $1/v$ law:
\begin{equation}
\sigma_a(E_n)
=
5333
\sqrt{
\frac{25.3~\mathrm{meV}}{E_n}
}
~\mathrm{barn}.
\label{eq:he3_energy_dependence}
\end{equation}
Owing to the large spin dependent absorption cross section, the polarized $^{3}$He gas selectively absorbs neutrons with spin antiparallel to the $^{3}$He nuclear spin, thereby producing a polarized neutron beam in transmission with a small scattered-neutron background.

Let $P_{\mathrm{He}}$ be the polarization of the $^{3}$He gas, $\rho$ the number density of $^{3}$He atoms, and $d$ the length of the $^{3}$He gas. The neutron transmission $T_n$ and neutron polarization $P_n$ after passing through the spin filter are expressed as
\begin{eqnarray}
T_n &=& T_{\mathrm{cell}} 
       \exp\!\left(-\rho d \sigma_0(E_n)\right)
       \cosh\!\left(P_{\mathrm{He}} \rho d \sigma_p(E_n)\right)\label{Tn}, \\
P_n &=& \tanh\!\left(P_{\mathrm{He}} \rho d \sigma_p(E_n)\right)\label{Pn},
\end{eqnarray}
where $T_{\mathrm{cell}}$ is the neutron transmission of the glass cell itself and is approximately 0.9 in this work.

\begin{figure}[!h]
\centering\includegraphics[width=10cm]{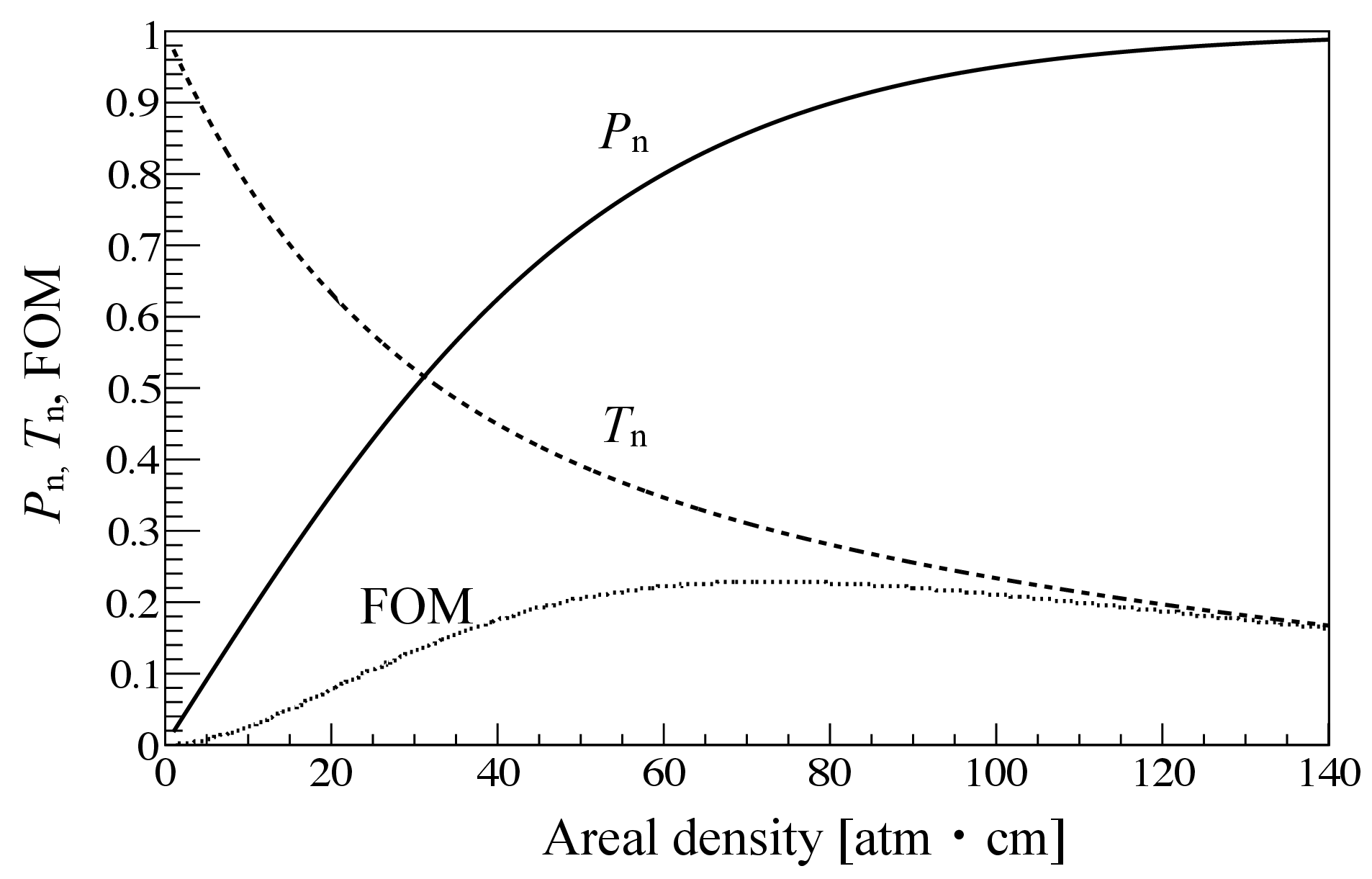}
\caption{
Calculated neutron polarization $P_n$, neutron transmission $T_n$, and figure of merit $\mathrm{FOM}=P_n^2T_n$ as functions of the $^{3}$He areal density $\rho d$ for 0.75~eV neutrons. The calculation assumes a $^{3}$He polarization of $P_{\mathrm{He}}=70\%$. The solid, dashed, and dotted curves represent $P_n$, $T_n$, and the FOM, respectively.}
\textbf{Alt text:} Plot showing calculated neutron polarization, neutron transmission, and figure of merit as functions of $^{3}$He areal density. The neutron polarization increases with areal density, the transmission decreases, and the figure of merit has a maximum near 74~atm$\cdot$cm.
\label{FOM}
\end{figure}

To maximize the statistical sensitivity of spin-dependent measurements, a figure of merit (FOM) can be defined as
\begin{equation}
\mathrm{FOM} = P_n^2 T_n .
\end{equation}
Because the absorption cross section depends on neutron energy, the FOM exhibits a maximum at a specific areal density $\rho d$ for a given energy. For neutrons with an energy of 0.75~eV, the FOM is maximized at an areal density of approximately 74~atm$\cdot$cm, as shown in Fig.~\ref{FOM}. Here, the $^{3}$He number density is expressed in units of atm; in this convention, 1~atm corresponds to a number density of $2.69\times10^{19}~\mathrm{cm}^{-3}$ at 0~$^\circ$C. 
In the present study, a spin filter with an effective $^{3}$He gas length of 19.6~cm and a pressure of 3.2~atm, was chosen, corresponding to an areal density of 63~atm$\cdot$cm. This length, i.e., the length of the glass cell, was determined by the inner dimensions of the baking heater used during the cell fabrication process, as described below.

\section{Construction of the $^{3}$He spin filter system for epithermal neutrons}
\subsection{Fabrication of the $^{3}$He gas cell}
The $^{3}$He gas cell was fabricated with the filling station at J-PARC using the procedure described in Ref.~\cite{He3Okudaira}. A boron-free aluminosilicate glass cell (GE180) was used because of its small wall-relaxation effect and low $^{3}$He permeability. The cell contained $^{3}$He gas, a small amount of N$_2$ buffer gas, and alkali metals (Rb and K) for hybrid spin-exchange optical pumping~\cite{babcock2003}.

\begin{figure}[!h]
\centering\includegraphics[width=10cm]{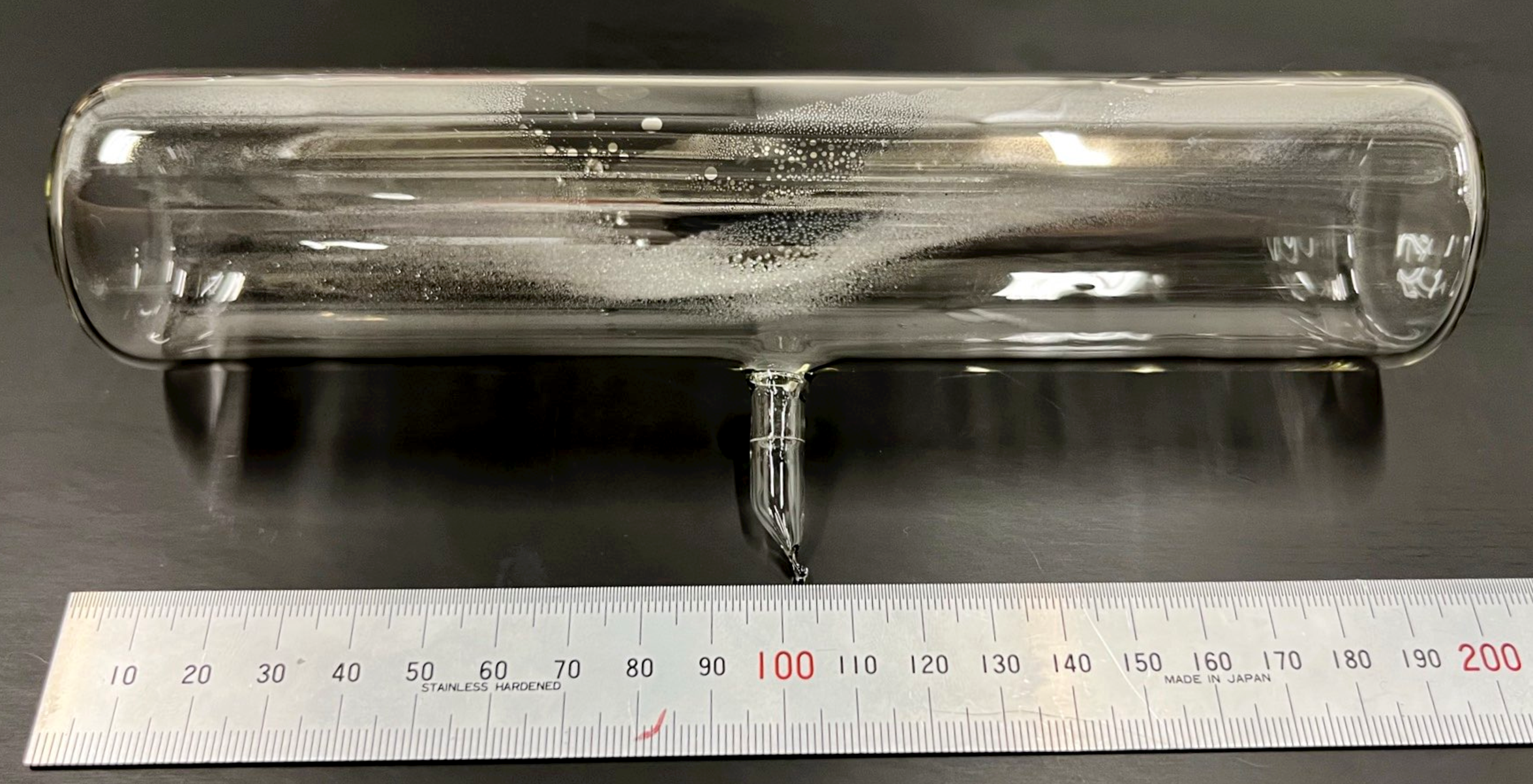}
\caption{Photograph of the fabricated $^{3}$He gas cell. The cell has an outer diameter of 45~mm and a length of 200~mm.}
\label{Hecell}
\textbf{Alt text:} Photograph of the fabricated cylindrical $^{3}$He gas cell with a ruler. The cell is approximately 45~mm in diameter and 200~mm in length.
\end{figure}

After cleaning and baking the glassware at approximately 400~$^\circ$C for about two weeks, the ultimate pressure reached approximately $1\times10^{-7}$~Pa. Rb and K were then introduced to the glass cell by vacuum distillation, followed by the addition of N$_2$ and $^{3}$He gas. The cell was sealed using a gas torch at a $^{3}$He pressure of 95~kPa at liquid-nitrogen temperature. The gas pressures at room temperature were estimated to be approximately 100~Torr for N$_2$ and 3.2~atm for $^{3}$He. A photograph of the fabricated cell is shown in Fig.~\ref{Hecell}. The outer dimensions of the cell are 45~mm in diameter and 200~mm in length, with a glass thickness of approximately 2~mm. 
The Rb/K number-density ratio was estimated to be approximately 15 at 230~$^\circ$C from white-light transmission measurements, following the method described in Appendix~A of Ref.~\cite{Takahashi2025}.

\subsection{Laser optical system for spin-exchange optical pumping}
In order to polarize the $^{3}$He gas, circularly polarized laser light was irradiated onto the fabricated $^{3}$He cell at an operating temperature of 200--230~$^\circ$C to optically pump the Rb atoms.  During collisions between polarized Rb/K atoms and $^{3}$He nuclei, spin polarization is transferred to the $^{3}$He nuclei through the hyperfine interaction. Because the pumping laser is attenuated by Rb absorption over the 20-cm cell length, the optical system was designed to irradiate circularly polarized laser beams into the cell from both ends.

\begin{figure*}[t]
\centering\includegraphics[width=17cm]{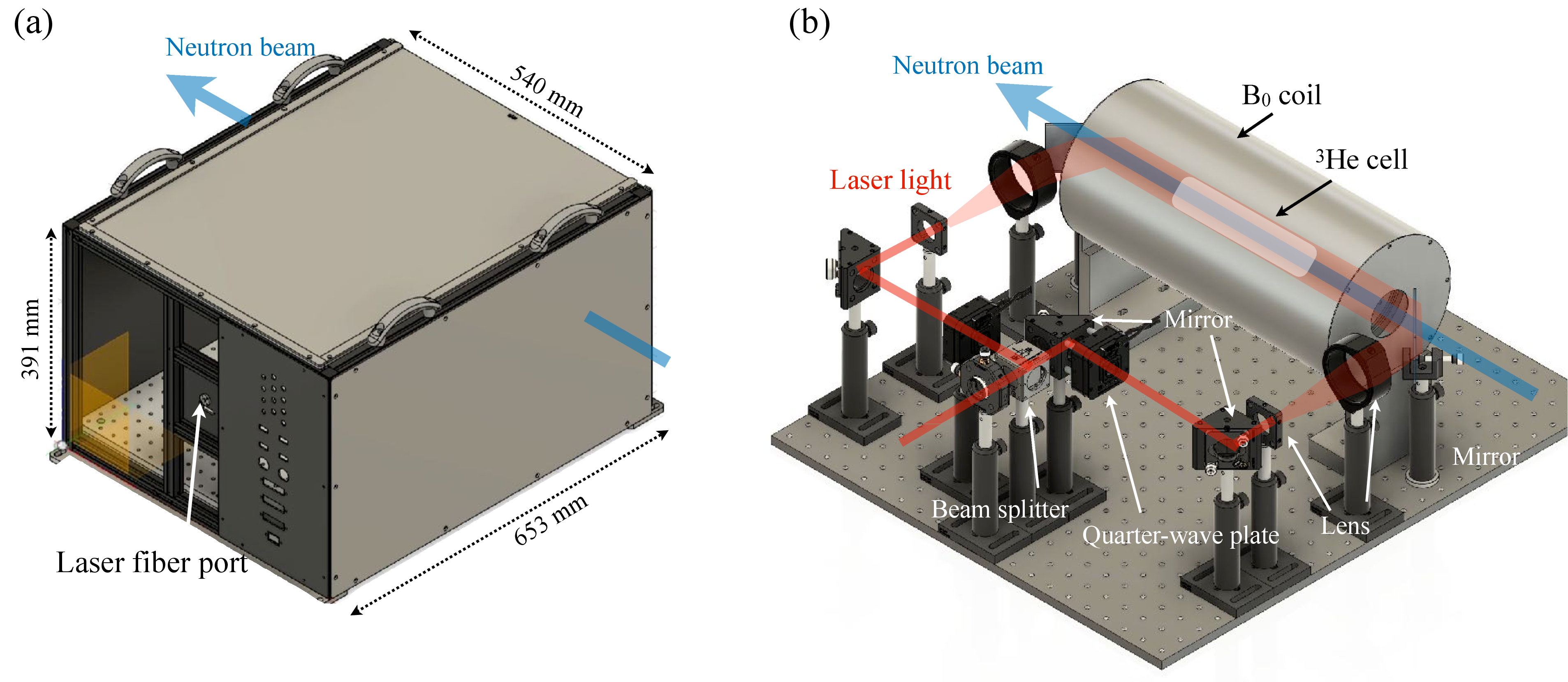}
\caption{Laser optics system for in-situ SEOP. (a) Outer view. (b) Inside view.}
\textbf{Alt text:} Schematic drawings of the compact laser-optics system for in-situ SEOP. The system includes the laser-fiber input, beam splitter, quarter-wave plates, mirrors, lenses, and the $^{3}$He cell.
\label{Optics}
\end{figure*}

A schematic view of the laser optical system is shown in Fig.~\ref{Optics}. The overall dimensions of the laser shielding box are 540~mm $\times$ 653~mm $\times$ 391~mm, allowing installation at the Accurate Neutron-Nucleus Reaction Measurement Instrument (ANNRI, BL04) in the Materials and Life Science Experimental Facility (MLF) at J-PARC without interference with other experimental components. The laser source was connected to the shielding box through an optical fiber.  The pumping laser was a fiber-coupled diode laser (DILAS Diode Laser, Inc.; M1F4S22-794.7.[0.6]-100C-IS9) with a maximum output power of 110~W. The center wavelength was tuned to 794.7~nm, corresponding to the D1 transition line of Rb, with a full width at half maximum of 0.4~nm using a volume Bragg grating. The laser light was delivered through an optical fiber with a core diameter of 400~$\mu$m, collimated by a lens, and split into two optical paths using a polarizing beam splitter. In each path, the linearly polarized light was converted into circularly polarized light using a quarter-wave plate and then guided to the $^{3}$He cell by dielectric mirrors and lenses.  The quarter-wave plates can be rotated in synchronization with the spin flips of $^{3}$He, as described below, so that the handedness of the circular polarization is reversed together with the $^{3}$He spin direction. The degree of circular polarization at the cell position was measured with a polarimeter and exceeded 98\% for both optical paths. Along the neutron beam path, the neutron beam passed through two 2-mm-thick aluminum plates, two 1-mm-thick quartz laser windows installed inside the B$_0$ coil, and two 2.3-mm-thick quartz mirrors tilted by 45$^\circ$ with respect to the neutron beam, corresponding to an effective neutron path length of 3.3~mm per mirror. The neutron transmission of the entire laser system, excluding the $^{3}$He cell, was measured to be approximately 0.8.

\subsection{Magnetic field design and coil system}

\begin{figure*}[t]
\centering\includegraphics[width=17cm]{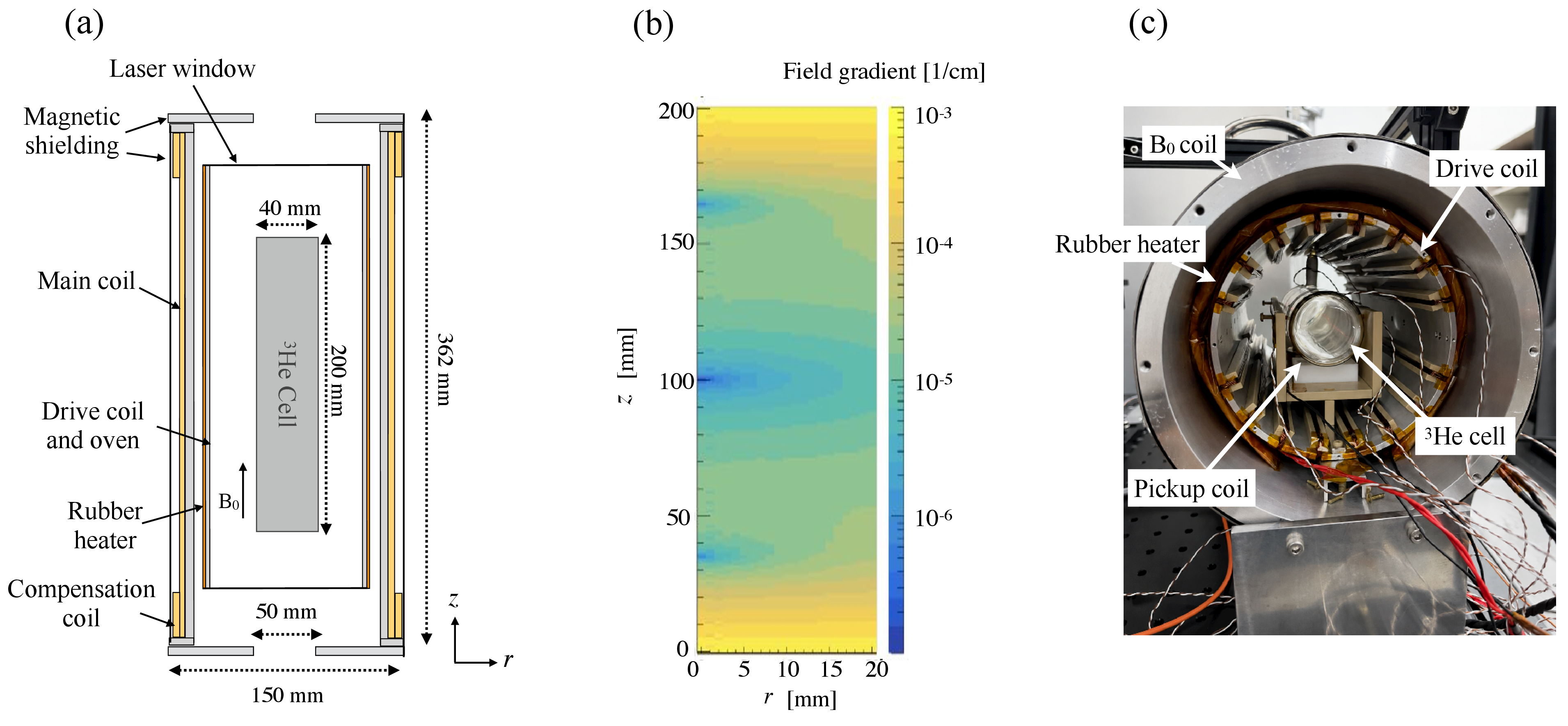}
\caption{(a) Cross sectional view of the coil system. (b) The calculation result of the magnetic field gradient in cylindrical coordinate system. (c) Interior of the coil. The lids of the magnetic shielding and laser windows were removed.}
\label{coil}
\textbf{Alt text:} Cross-sectional drawing, calculated field-gradient map, and photograph of the magnetic-field coil system. The figure shows the arrangement of the main coil, compensation coil, drive coil, oven, magnetic shielding, and $^{3}$He cell.
\end{figure*}

The polarization of $^{3}$He gas is sensitive to magnetic field inhomogeneities, which induce spin relaxation.
Therefore, a uniform magnetic field is required to suppress depolarization and maintain high $^{3}$He polarization.
The relaxation rate caused by the magnetic field gradient is described as 
\begin{eqnarray}
\Gamma_B=\frac{6700}{p_{\rm{He}}}\frac{|\nabla B_x|^2+|\nabla B_y|^2}{B_0^2}~[\rm{hour}^{-1}]
\end{eqnarray}
where $p_{\rm{He}}$ is the $^3$He gas pressure in bar, $B_0$ is the primary magnetic field, and $B_x$, $B_y$ are the transverse magnetic fields with respect to the primary magnetic field~\cite{Cates1988}. The normalized transverse-field gradient, $\sqrt{|\nabla B_x|^2+|\nabla B_y|^2}/B_0$, is expressed in units of cm$^{-1}$. The magnetic field configuration was optimized using finite-element simulations to determine the coil diameter, the number of turns of the main coil, the compensation coil parameters, and the magnetic shielding~\cite{Asai2026}. The detailed dimensions of the coil system are shown in Fig.~\ref{coil}.
The resulting coil system provides the normalized transverse-field gradient less than $10^{-3}$~cm$^{-1}$ throughout the volume occupied by the $^{3}$He cell. The relaxation time associated with magnetic field inhomogeneity was calculated as $10^4$ hours, which is much longer than the relaxation time expected from dipole--dipole interactions between $^{3}$He atoms.  
With a static magnetic field $B_0$ of 1.5~mT applied, the longitudinal relaxation time of the $^{3}$He polarization was measured to be approximately 50~h. This relaxation is considered to arise mainly from the wall relaxation, and impurity-induced relaxation inside the cell. This relaxation time is sufficiently longer than the polarization build-up time discussed later, allowing a high equilibrium polarization to be achieved. \\
A rubber heater of 200~W was installed inside the coil together with a drive coil described below, enabling the cell temperature to be raised to 200--230~$^\circ$C during laser irradiation. The cell temperature was monitored using the thermocouples attached to the cell surface. The oven lid has the quartz window with anti-reflective coatings for the incident laser light.

\subsection{Spin-flip system}

An important advantage of the $^{3}$He spin filter is that the neutron spin direction can be reversed by flipping the $^{3}$He nuclear spin.
The spin-flip system used in this study is based on the adiabatic fast passage nuclear magnetic resonance (AFP-NMR) technique described in Ref.~\cite{ino2017}, which enables both reversal of the $^{3}$He spin direction and monitoring of the $^{3}$He polarization with a low polarization loss.
An oscillating magnetic field was applied using a cosine-theta coil installed inside the $B_0$ coil.
With a static magnetic field of 1.5~mT, corresponding to a Larmor frequency of approximately 50~kHz, the RF frequency was swept from 40 to 60~kHz within 0.2~s to induce spin reversal. 
A pickup coil installed under the cell was used to detect the spin precession signal, allowing confirmation of the spin flip and monitoring the $^{3}$He polarization. The polarization loss per spin flip was evaluated from the spin-precession signal and was found to be approximately $2\times10^{-4}$.

\section{Beam experiment}
The $^{3}$He spin filter system was installed at the ANNRI beamline at J-PARC. It was used in two measurements: a P-violation measurement with an unpolarized La target and a spin-correlation measurement with a polarized La target. In the latter measurement, the La target was statically polarized at 0.3~K in a 4~T magnetic field.

\begin{figure*}[t]
\centering\includegraphics[width=13cm]{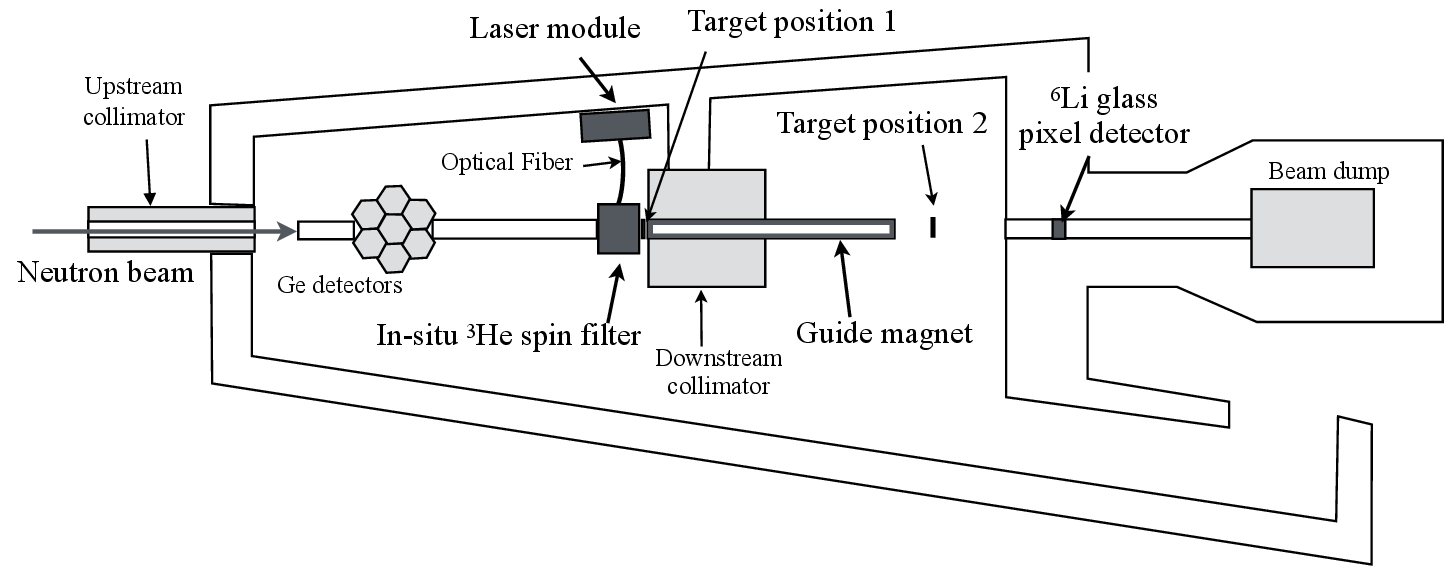}
\caption{Top view of the experimental setup at beamline04.}
\label{Beamline}
\textbf{Alt text:} Schematic top view of the experimental setup at BL04 ANNRI. The in-situ $^{3}$He neutron spin filter, target positions, collimators, guide magnet, germanium detectors, pixel detector, and beam dump are shown along the neutron beamline.
\end{figure*}

A schematic view of the experimental setup is shown in Fig.~\ref{Beamline}.
The spin filter was installed upstream of the downstream collimator, and a pixelated neutron detector based on a $^{6}$Li-glass scintillator (LiTA) \cite{Liglass2015} was placed 30.26~m downstream from the moderator surface to measure the transmitted neutron beam. In the P-violation measurement, the La sample was placed at target position 1. In the spin-correlation measurement, the polarized La target, together with the 0.3~K refrigerator and the 4~T superconducting magnet, was installed at target position 2. The distance between $^3$He spin filter and target potision 2 was approximately 5~m.

At J-PARC, the proton-beam pulses were injected to a neutron production target at a repetition rate of 25 Hz and an average beam power of 750 kW during the measurement. The incident neutron energy is determined with the time of flight (TOF) from the proton injection signal to the neutron production target as the time origin. The neutron energy is related to TOF by
\begin{equation}
E_n({\rm{TOF}})
=
\frac{1}{2}m_n
\left(
\frac{L}{{\rm{TOF}}}
\right)^2,
\label{eq:tof_energy}
\end{equation}
where $m_n$ is the neutron mass and $L$ is the flight path length.

To determine the areal density of the $^{3}$He gas, the neutron transmission spectrum measured with the unpolarized $^{3}$He cell was divided by that measured with the empty cell. The TOF-dependent transmission ratio was fitted with an exponential attenuation function described as  $\exp\!\left(-\rho d \sigma_0(E_n)\right)$ with a fitting parameter of $\rho d$, as shown in Fig~\ref{Tn}.
The resulting areal density was $\rho d =64.0\pm 0.4$~atm$\cdot$cm, which is reasonably consistent with the nominal design value.

\begin{figure}[!h]
\centering\includegraphics[width=10cm]{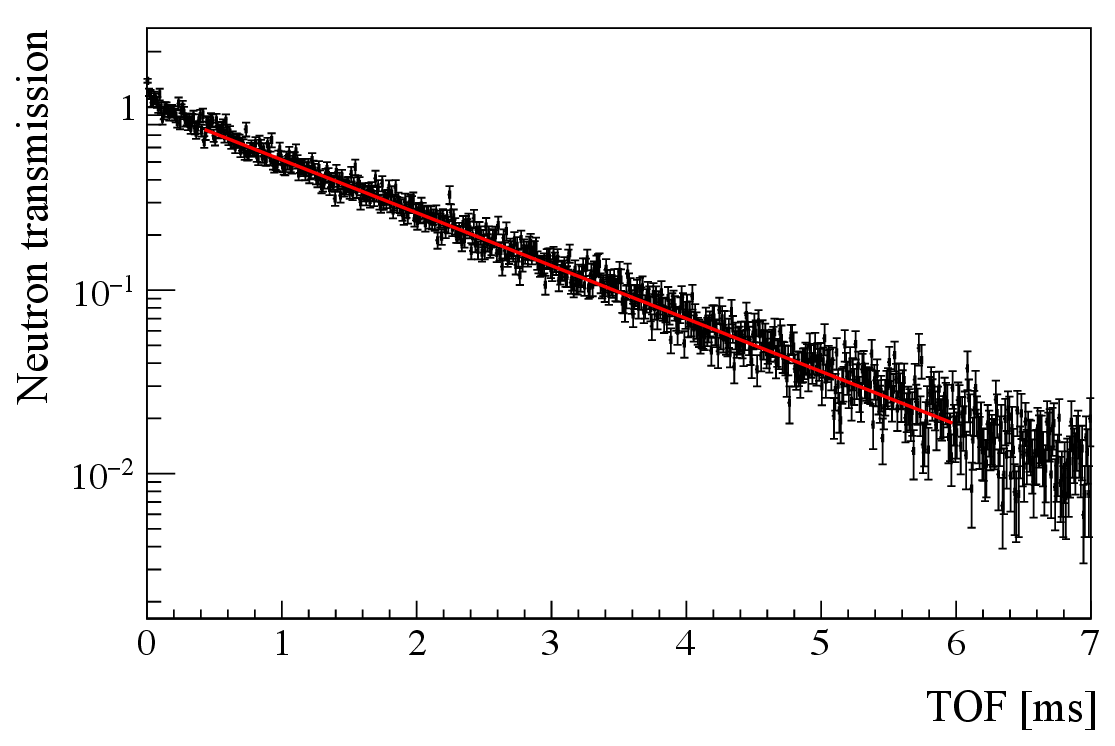}
\caption{The neutron transmission spectrum for unpolarized $^3$He cell. The red line shows the best fit.}
\label{Tn}
\textbf{Alt text:} Plot of the neutron transmission spectrum for the unpolarized $^{3}$He cell as a function of time of flight. The measured transmission decreases with time of flight and is fitted by an exponential attenuation curve.
\end{figure}

\begin{figure}[!h]
\centering\includegraphics[width=10cm]{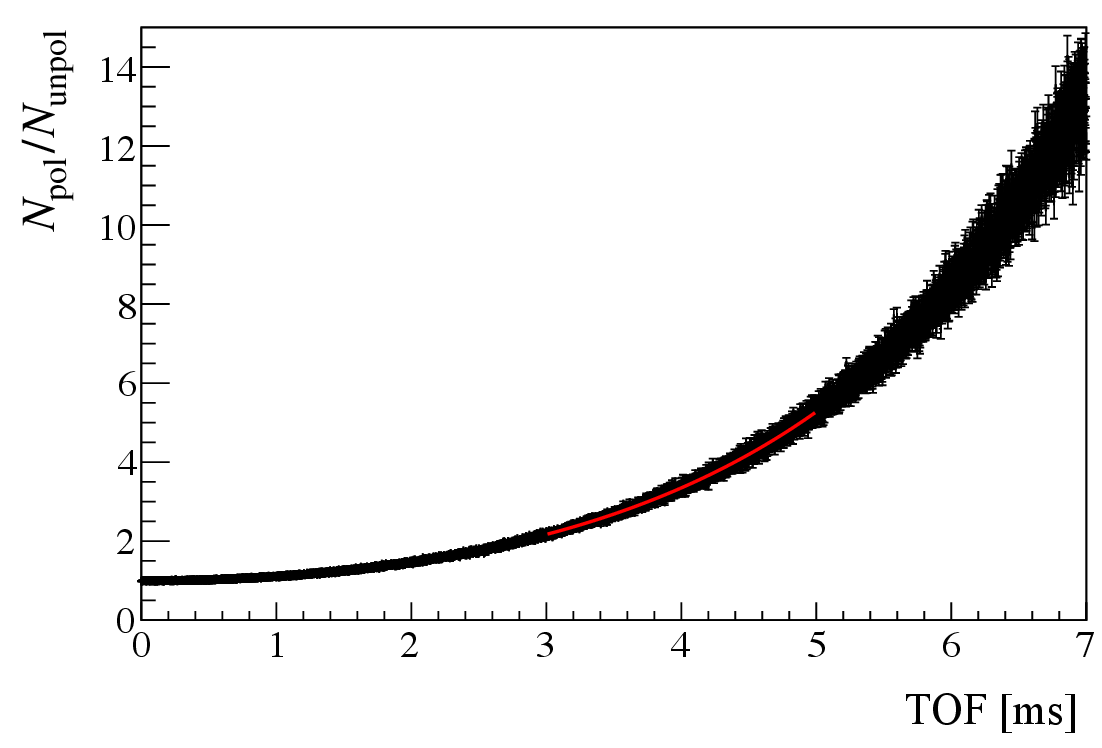}
\caption{Ratio of the numbers of transmitted neutrons for the polarized to the unpolarized $^3$He cell. The red
line shows the best fit in the 3--5~ms.}
\textbf{Alt text:} Plot of the ratio of transmitted neutrons for the polarized and unpolarized $^{3}$He cell as a function of time of flight. The ratio increases with time of flight and is fitted in the 3--5~ms region.
\label{cosh}
\end{figure}

The $^{3}$He gas was polarized by irradiating the cell with laser light at a temperature of 230~$^\circ$C, and the neutron transmission was measured during the experiment to determine the neutron and $^3$He polarizations. 
 The ratio of the transmitted neutrons for polarized $^3$He cell $N_{\rm{pol}}$ and unpolarized $^3$He cell $N_{\rm{unpol}}$ can be described by using Eq.~\ref{Tn} as 
\begin{eqnarray}
\frac{N_{\rm{pol}}}{N_{\rm{unpol}}}= \cosh\!\left(P_{\mathrm{He}} \rho d \sigma_p(E_n)\right).\label{ratio}
\end{eqnarray}
The product of the parameters $P_{\mathrm{He}} \rho d$ was determined by fitting the TOF dependence of $N_{\rm{pol}}/N_{\rm{unpol}}$ using Eq.~{\ref{ratio}} as shown in Fig.~\ref{cosh}. The TOF-dependent $\gamma$-ray background originating from upstream components was evaluated independently by inserting resonance filters into the beam and was subtracted from the measured TOF spectra before forming the ratio.
When the fitting range was varied within 2.5--7~ms, the fitted value of $P_{\mathrm{He}}\rho d$ varied by 0.5\%. This variation was taken as a systematic uncertainty. This fit-range dependence may be due to residual $\gamma$-ray background generated along the beamline and/or scattered-neutron background from materials in the beam path, which were not fully removed by the upstream-background subtraction. 

\begin{figure*}[t]
\centering\includegraphics[width=13cm]{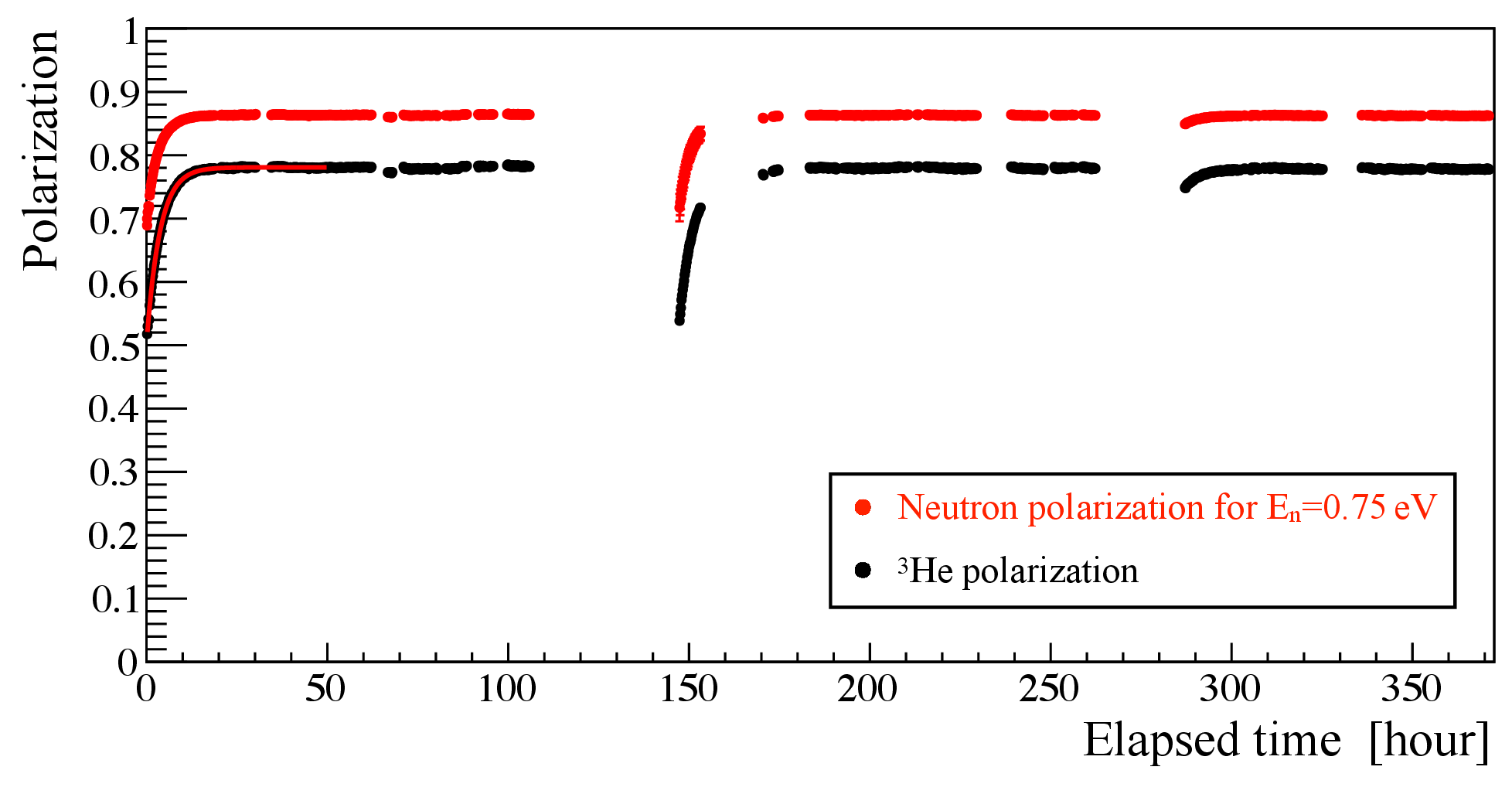}
\caption{$^3$He polarization (black point) and corresponding neutron polarization for $E_n=0.75~\rm{eV}$ (red point) during the neutron beam experiment. The red line from 0 to 50 hours shows the best fit to obtain the build up time of the $^3$He polarization. The intervals from 100 to 140 h and from 270 to 280 h correspond to measurements with an unpolarized neutron beam. For this purpose, the $^{3}$He gas was intentionally depolarized and was repolarized after each measurement period. The remaining gaps in the data correspond to periods when the neutron beam was unavailable because of accelerator interruptions.}
\label{PHe_time}
\textbf{Alt text:} Time evolution of the $^{3}$He polarization and the corresponding neutron polarization at 0.75~eV during the beam experiment. The polarization builds up rapidly and remains stable over repeated measurement periods.
\end{figure*}

During the measurement, the $^{3}$He spin direction was reversed every 20~min. 
Figure~\ref{PHe_time} shows the time evolution of the $^{3}$He polarization measured by the neutron transmission.
The polarization build-up time was obtained by fitting the time evolution of the $^{3}$He polarization with
\begin{equation}
P_{\mathrm{He}}(t)=P_{\mathrm{He}}^{\infty}\left(1-\exp\left(-\frac{t}{\tau}\right)\right),
\end{equation}
where $P_{\mathrm{He}}^{\infty}$ is the equilibrium polarization of $^3$He and $\tau$ is the build-up time constant. The obtained value of $\tau$ was approximately 3.7~hours. The final equilibrium $^{3}$He polarization reached 78.1$\pm 0.6$\%.  During the experiment, with a total measurement time of 370~hours, the fluctuation of the $^3$He polarization was within the error bar.\\
The neutron polarization was calculated using Eq.~\ref{Pn} and fitted value of  $P_{\mathrm{He}} \rho d$. The maximum neutron polarization at $E_n=0.75~\mathrm{eV}$ was $86.4\pm0.5\%$.

\begin{figure}[!h]
\centering\includegraphics[width=10cm]{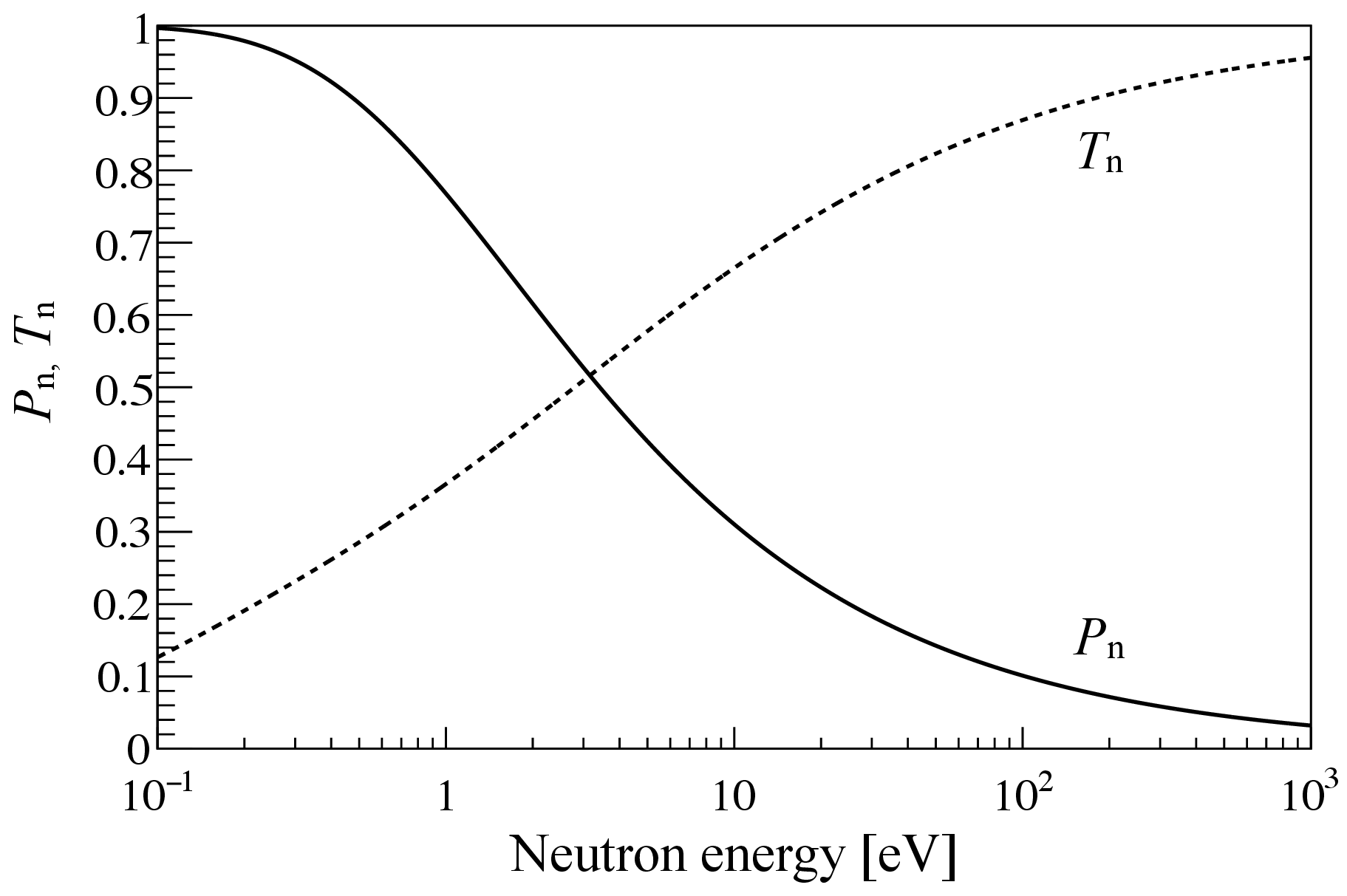}
\caption{Energy dependence of the neutron polarization (solid line) and neutron transmission (dashed line) for the 64~atm$\cdot$cm cell with $P_{\rm {He}}$=78\% }
\label{PnTn}
\textbf{Alt text:} Plot of neutron polarization and neutron transmission as functions of neutron energy for the 64~atm\,cm $^{3}$He spin filter. The neutron polarization is high below 1~eV and decreases with increasing neutron energy, while the transmission increases.
\end{figure}

The energy dependence of the neutron polarization and transmission with this condition is also shown in Fig.~\ref{PnTn}. 
For reference, $P_n(1\,\mathrm{eV}) = 80\%$, 
$P_n(10\,\mathrm{eV}) = 34\%$, and 
$P_n(100\,\mathrm{eV}) = 11\%$. 
These results demonstrate that highly polarized neutrons were produced below 1~eV, while practically useful neutron polarization was also achieved at higher neutron energies.


\section{Conclusion}

We have developed and demonstrated a highly polarized epithermal neutron beam using a compact in-situ polarized $^{3}$He spin filter system at J-PARC, a world-leading high-intensity pulsed neutron facility. A 20-cm-long, 3.2-atm $^{3}$He gas cell was fabricated and polarized using the hybrid SEOP method. The spin filter system was successfully operated on the BL04 ANNRI beamline, where polarized neutrons were produced and characterized using time-of-flight measurements. A $^{3}$He polarization of 78.1$\pm 0.6$\% was achieved, resulting in a neutron polarization of 86.4$\pm$0.5\% at a neutron energy of 0.75~eV. This value represents, to our knowledge, the highest neutron polarization reported to date for epithermal neutrons in the 1~eV energy region. The polarized state was stably maintained for more than 100~h, which is sufficient for practical beam experiments at pulsed neutron sources. The polarized-neutron beam demonstrated in this work was used for P-violation and spin-correlation measurements at J-PARC. The physics results obtained from these measurements will be reported in separate publications. This $^3$He spin filter system is directly applicable to ongoing and planned experiments searching for T-violating interactions in NOPTREX. 

The present system provides highly polarized epithermal neutrons over a broad energy range and may therefore be useful for various applications. In addition to the $^{139}$La measurements discussed above, it can be applied to symmetry-violation studies using other nuclei with enhanced parity violation in eV neutron resonances, such as $^{81}$Br, $^{117}$Sn, and $^{131}$Xe~\cite{Mitchell1999}.
Possible applications also include polarized inelastic neutron scattering, magnetic-field imaging~\cite{Hiroi2018JPSCP}, and measurements of the spin-dependent properties of compound-nuclear states relevant to nuclear-data studies~\cite{Keyworth1973U}.

\  
\section*{Acknowledgements}
The authors would like to thank the staff of the MLF and the J-PARC for operating the accelerators and the neutron production target. The authors also thank the Technical Center, School of Science, Nagoya University, for technical support in the development and construction of the spin-filter system. The neutron experiment at the MLF of the J-PARC was performed under user programs Nos. 2025A0264, 2025A0357 and 2026A0257.  This work was supported by MEXT, Japan KAKENHI Grant Nos.
20K14495, 23K13122, 24H00222, and 25K01024.


%

\vspace{0.2cm}
\noindent


\let\doi\relax

\bibliographystyle{ptephy}
\bibliography{Author_tex}

\end{document}